\documentstyle[11pt,adassconf]{article}  

\begin{document}   

%
%
%

\paperID{P3.4}

%
%

\title{XAssist: A System for the Automation of X-ray Astrophysics Analysis}

%
%
%

\author{A. Ptak}
\affil{Johns Hopkins University}

\author{R. Griffiths}
\affil{Carnegie Mellon University}

%
%

\contact{Andy Ptak}
\email{ptak@pha.jhu.edu}

%
%
%

\paindex{Ptak, A.}
\aindex{Griffiths, R. }     

%
%



\begin{abstract}          
XAssist is a NASA AISR-funded project for the automation of X-ray
astrophysics, with emphasis on galaxies.  It is nearing completion of
its initially funded effort, and is working well for {\it Chandra} and
{\it ROSAT} 
data.  Initial support for {\it XMM-Newton} data is present as well.
It is capable of data reprocessing, 
source detection, and preliminary 
spatial, temporal and spectral analysis for each source with
sufficient counts.  The bulk of the system is written in Python, which
in turn drives underlying software (CIAO for {\it Chandra} data,
etc.).  Future work will include a GUI (mainly for beginners and
status monitoring) and the exposure of at least some functionality as
web services.  The latter will help XAssist to eventually become part
of the VO, making advanced queries possible, such as determining the
X-ray fluxes of counterparts to HST or SDSS sources (including the use
of unpublished X-ray data), and add the ability of ``on-the-fly'' X-ray
processing.  Pipelines are running on {\it ROSAT}, {\it Chandra} and now
{\it XMM-Newton} observations of galaxies to demonstrate XAssist's
capabilities, and the results are available online (in real time) at
\htmladdnormallink{http://www.xassist.org}{http://www.xassist.org}.
XAssist itself as well as various associated
projects are available for download.
\end{abstract}

%
%

\section{Introduction}
We are currently in a renaissance for X-ray astronomy, with two major
missions, {\it Chandra} and {\it XMM-Newton}, currently operating.
These missions are producing large amounts of archival data, which is
supplementing 
existing databases from missions such as {\it ROSAT} and {\it ASCA}.
Historically, only X-ray ``experts'' usually attempted the analysis of
X-ray data.  This is because there were fundamental differences in the
analysis of X-ray data compared with other bandpasses, most notably
the fact that individual photons are detected as opposed to the
accumulation of (only) spectra or images.  Most modern X-ray detectors are
imaging spectrometers so each observation results in a photon list
from which images, spectra and light curves can be extracted.  In
general the numbers of photons are small so Poissonian rather
than Gaussian statistical methods must be used.  The spectral and
spatial resolution of most detectors is moderate at best and
forward-fitting convolution methods are needed to properly fit the
data.  All of these factors limit the accessibility of the X-ray data
to non-experts.  In addition each mission tends to
have its own unique software package for the reduction and analysis of
the data, and X-ray data often require reprocessing as the calibration
improves.  These latter two factors also limit the ability of experts
to take advantage of all available data for a given project,
particularly large-scale surveys.

We have developed a software package to address these concerns.
XAssist is capable of
performing data reduction and preliminary 
analysis for {\it ROSAT}, {\it Chandra} and {\it XMM-Newton} data.  It is fully
automatic making it well-suited for surveys, as well as for the
reprocessing of existing data.  Below we will discuss its
capabilities and prospects for the future.

\section{Capabilities}
XAssist currently has the following capabilities:
\begin{itemize}
\item Downloads data
\item Reprocesses data
\item Creates exposure maps and detector masks (if possible)
\item Detects sources (using built-in routine for {\it ROSAT} and {\it ASCA}, CIAO
  wavdetect for {\it Chandra}, etc.) 
\item Fits each source with "simple" (i.e., not including point-spread
  function) model to establish source extent and (Poisson-correct)
  significance (using the stand-alone python program
  \htmladdnormallinkfoot{ximgfit}{http://www.xassist.org/xassist/Download.jsp}) 
\item Flags extended, confused and problematic sources
\item Computes median (or mean) background level
\item Excludes times of high background
\item Extracts spectra, ``postage stamp'' images, and light curves of
  each source for more detailed analysis (a simple power-law model is
  fit to sources with more than 100 sources)
\item Optionally performs chip-by-chip analysis (relevant just for
  {\it Chandra} right now) 
\item Analysis can be restricted to an energy band
\item Large emphasis on detailed reporting
\item Looks for correlations of X-ray sources with astronomy databases
  and provides links on the detailed source web reports to query
  Simbad (see below)
\end{itemize}

Note that in the case of {\it Chandra} analysis, most of the data reduction
steps are based on the ``threads'' reported in the
\htmladdnormallinkfoot{{\it Chandra} web site}{http://cxc.harvard.edu}.
Most parameters controlling XAssist are read from IRAF/FTOOLS-style
parameter file, and can be set on the command-line as well (allowing
for the automated setup of XAssist for surveys).  XAssist
can be run (and configured) interactively (with a text-based interface).

\section{Sample Output}
\begin{figure}[htbn]
\epsscale{0.75}
\plottwo{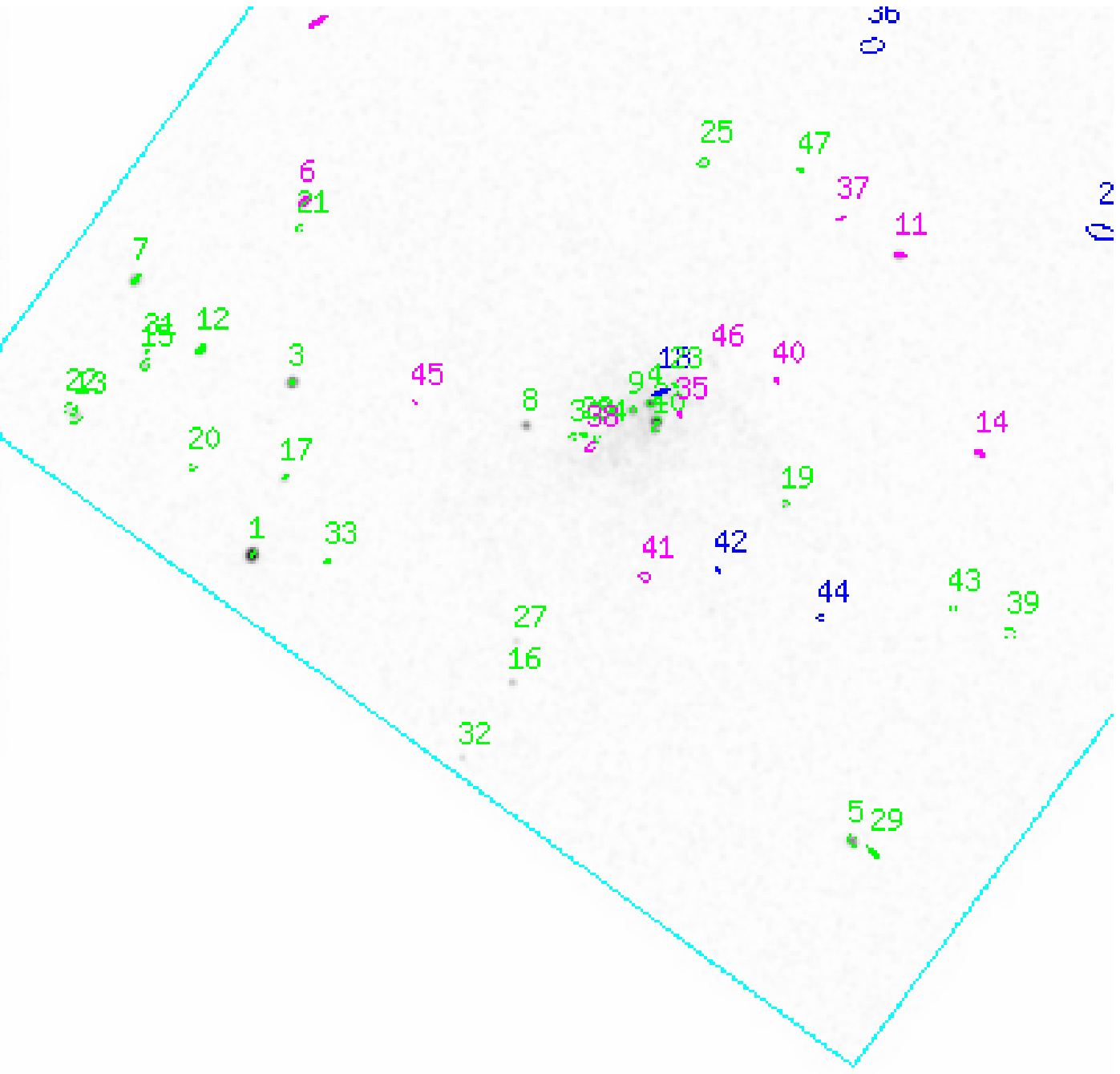}{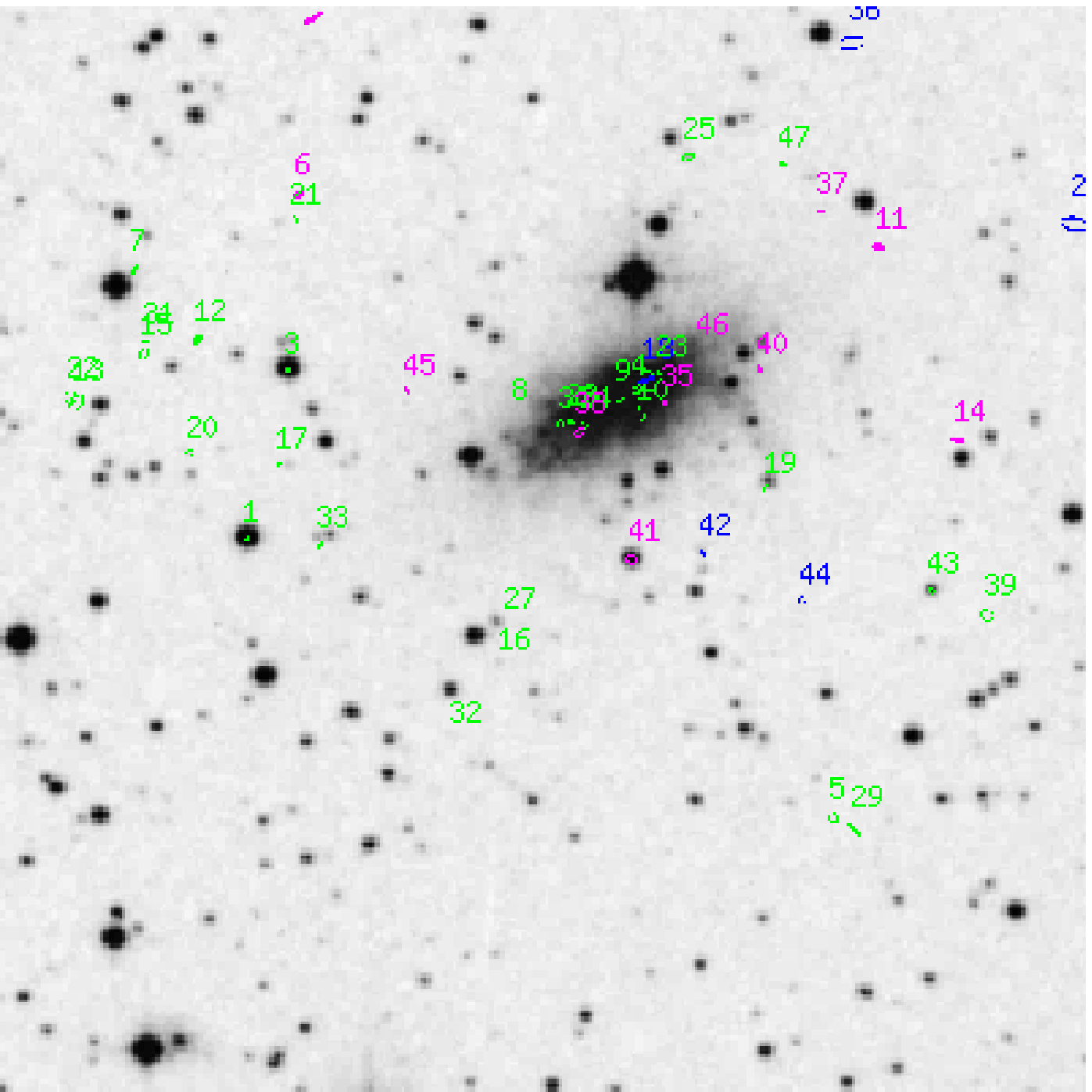}
\caption{{\it Chandra} ACIS (left) and DSS (right) images of NGC 1569.
  Detected X-ray sources are marked in both images.  The color code
  is:green = point source, red = problematic/questionable source, blue
  = extended source, magenta = asymmetric source (may be extended),
  cyan = estimated detector boundary.\label{n1569_chandra_dss_fig}}
\end{figure}

\begin{figure}[htbn]
\epsscale{0.7}
\plotone{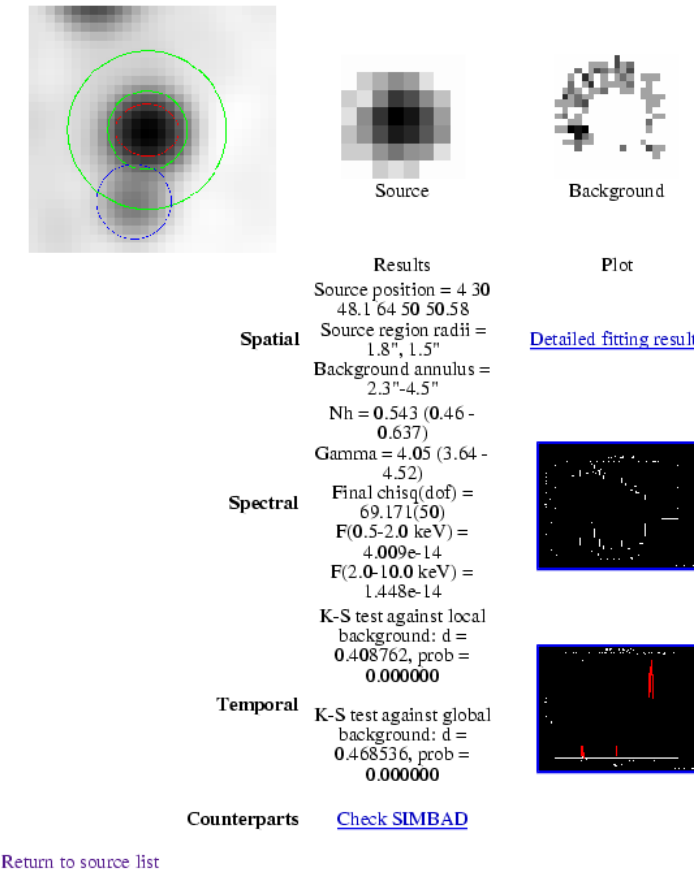}
\caption{Detailed web page report for a source in the {\it Chandra}
  observation of NGC 1569. \label{n1569_chandra_src2_fig}}
\end{figure}

Figure \ref{n1569_chandra_dss_fig} shows images created as part of the
report for the processing of the {\it Chandra} observation NGC 1569.  Figure
\ref{n1569_chandra_src2_fig} shows the detailed report generated for a
source.  While there are admittedly ``warts'' that occur (as in any
automated system), this example demonstrates that even in moderately
crowded fields the system performs well and continues to the point of
fitting a power-law spectrum model to the source spectrum.  Obviously,
this opens a powerful possibility for virtual observatories, namely that
searches could be performed on high-level quantites such as spectral
slope.  Even though human inspection would of course still be
necessary for science-grade results, an automated system such as this
could cull samples and produce useable results for many of the
sources, both of which would be particularly useful prior to observing
proposal deadlines (especially if the data of interest had only
recently become publically available and had not been published yet).
Queries are also submitted to HEASARC to find correlations of X-ray sources
with 2$\mu$ass, USNO, FIRST, and other catalogs.

\section{Pipelines}
To demonstrate XAssist's capabilities,
\htmladdnormallinkfoot{pipelines}{http://www.xassist.org/xassist/Pipelines.jsp}
have been set up to run XAssist on {\it ROSAT} HRI and {\it Chandra} observations
of RC3 galaxies.  The {\it ROSAT} analysis is nearing completion, and
crontab jobs are checking for the public release of {\it Chandra} observations
that may containing RC3 galaxies in the FOV of view, and these data sets are
downloaded (using the script cda.py also available for download).
{\it XMM-Newton} has recently been added as a supported mission, and we are
in the process of establishing a pipeline for public {\it XMM} data.  The
pipeline products are searchable and work is in progress to allow users to
request fields to be added to the processing lists.

\section{Current and Future Work}
\begin{itemize}
\item Finishing off details (especially for {\it ASCA} and {\it ROSAT} PSPC
  analysis)
\item Better installation and configuration support (including the
  creation of a GUI)
\item Web access to XAssist
\item Web services: limited status reporting is already available using
  SOAP and it will be possible to request full or partial processing
  of data via SOAP requests
\end{itemize}

\end{document}